# $\mathcal{PT}$-symmetric microring lasers: Self-adapting broadband mode-selective resonators

Hossein Hodaei, Mohammad-Ali Miri, Matthias Heinrich, Demetrios N. Christodoulides, and Mercedeh Khajavikhan[*]

*CREOL, College of Optics and Photonics, University of Central Florida, Orlando, Florida 32816–2700, USA*
[*]*mercedeh@creol.ucf.edu*

**Abstract**

We demonstrate experimentally that stable single longitudinal mode operation can be readily achieved in PT-symmetric arrangements of coupled microring resonators. Whereas any active resonator is in principle capable of displaying single-wavelength operation, selective breaking of PT-symmetry can be utilized to systematically enhance the maximum achievable gain of this mode, even if a large number of competing longitudinal or transverse resonator modes fall within the amplification bandwidth of the inhomogeneously broadened active medium. This concept is robust with respect to fabrication tolerances, and its mode selectivity is established without the need for additional components or specifically designed filters. Our results may pave the way for a new generation of versatile cavities lasing at a desired longitudinal resonance. Along these lines, traditionally highly multi-moded microring resonator configurations can be fashioned to suppress all but one longitudinal mode.

# 1. Introduction

The ability to control the modes oscillating within a laser resonator is of fundamental importance. The presence of competing modes can be highly detrimental to beam quality and spectral purity, thus leading to spatial as well as temporal fluctuations in the emitted radiation [1]. At first glance, one might expect these challenges to become less acute in the course of miniaturization, since the separation of resonances, or free spectral range, scales inversely with the system size. However, to achieve lasing action in small structures, comparably large amounts of gain have to be provided within a given resonator length. This requirement can be fulfilled by employing semiconductor quantum well systems [2] that offer amplification over a wide range of frequencies within their inhomogeneously broadened gain bandwidth. In such broadband gain environments, the lasing of the desired mode does not prevent the neighboring resonances from also experiencing amplification. Consequently, additional steps must be taken to suppress such parasitic modes. This can be accomplished in a number of ways, as for example coupling to detuned external cavities [3,4], by including intra-cavity dispersive elements such as DFB gratings [5,6], or by spatially modulating the pump [7]. However, not all of these techniques are practically compatible with every type of resonator, and each of them introduces further demands in terms of design complexity and fabrication tolerances. Clearly, of importance will be to identify alternative strategies for mode selection.

A prominent class of integrated laser arrangements is based on microring resonators. By virtue of their high refractive index contrast, such configurations can exhibit high quality factors and small footprints, thus making them excellent candidates for on-chip integrated photonic applications. However, like many other micro-scale resonators, these cavities tend to support multiple longitudinal modes with similar quality factors throughout their gain bandwidth – offering little control in terms of mode discrimination with conventional techniques.

Quite recently, the concept of parity-time (PT) symmetry was theoretically suggested as a means to achieve single-transverse-mode operation in both semiconductor and fiber lasers [8]. Since their introduction to the field of optics, PT-symmetric notions have attracted considerable attention [9-24]. In general, a non-Hermitian system is considered to be PT-symmetric, provided its associated Hamiltonian commutes with the parity-time (PT) operator. In that case, despite the presence of gain and loss, in the unbroken symmetry regime the entire spectrum of the system is real-valued. On the other hand, above a certain gain-loss contrast threshold, PT-symmetry can be spontaneously broken, in which case the spectrum becomes partially complex. In this regime, some of the modes experience either gain or loss, while the rest remains neutral.

In this article we show that PT symmetry breaking can be elegantly exploited to establish single-mode operation in inherently multi-moded micro-ring lasers. This is accomplished in a coupled arrangement of two structurally identical ring resonators, as long as one experiences gain, while the other one is subjected to an equal amount of loss. The key idea behind this mechanism is the fact that the threshold of symmetry breaking depends solely on the relation between gain/loss and coupling. Whereas any active resonator is in principle capable of displaying single-wavelength operation, selective breaking of PT-symmetry can be utilized to systematically enhance the maximum achievable gain of this mode, even if a large number of competing longitudinal or transverse resonator modes fall within the amplification bandwidth of the inhomogeneously broadened active medium. As a consequence, our approach naturally exhibits a type of broadband self-adaptive behavior and is, in principle, applicable to any active resonator configuration.

## 2. Theoretical model

### 2.1 Microring resonators

The small structure size associated with microring resonators necessitates waveguides with a high refractive index contrast. In order to avoid competition between modes of different polarization, the geometry of the waveguide cross section may be chosen so as to favor the TE component in terms of confinement and overlap with the active region. The interaction between two adjacent rings is mediated by the mutual overlap of their mode fields. In contrast to the continuous energy transfer between parallel waveguides, the ring curvature effectively confines the coupling to the narrow region of greatest proximity. Nevertheless, coherence is preserved by the periodic nature of this interaction, and the coupling coefficient $\kappa$ between rings can be determined from the resulting frequency splitting $\delta\omega$ between mode pairs.

To first order approximation, the modes of a ring resonator are defined by the optical path length along its circumference. As such, the free spectral range, or mode spacing, of a ring with radius $R$ and effective refractive index $n_{eff}$ is given by $\Delta\omega = c_0/Rn_{eff}$. Assuming an active medium with inhomogeneously broadened amplification profile $g(\omega)$, each and every mode whose gain exceeds the resonator losses can lase (see schematic in Fig. 1(a)). Given that the resonator size is generally limited by fabrication constraints as well as other design considerations, it may be impractical to decrease the structural dimensions until only a single resonance falls within the window of net gain.

### 2.2 Fundamentals of PT symmetry

In general, PT-symmetric structures involve gain and loss in a balanced fashion. Following the isomorphism between the respective evolution equations in optics and quantum mechanics, one can describe the distribution of amplification and attenuation through the imaginary part $n_I$ of a complex refractive index distribution $n(\vec{r}) = n_R(\vec{r}) + in_I(\vec{r})$. As previously indicated [9], this complex optical potential respects PT symmetry, provided that the real and imaginary parts exhibit even and odd spatial distributions, respectively. Perhaps the most fascinating feature of such systems is their ability to support eigenstates with entirely real eigenvalues, if PT symmetry is unbroken. Under these conditions, the modes neither decay nor grow, but rather remain neutral. On the other hand, once the gain-loss contrast is increased, this symmetry can be broken and as a result some of the modes may experience gain or loss in conjugate pairs.

### 2.3 PT-symmetric microring resonators

Let us first consider a pair of identical microring resonators, each of them supporting a number of modes throughout the amplification bandwidth. When placed in close proximity to one another, the coupling between the structures serves to break the degeneracy between their respective modes. The frequency splitting $\delta\omega$ of the resulting supermode doublets is determined by the mutual coupling coefficient $\kappa$. If both rings provide equal amounts of amplification, each of these modes can oscillate, thereby effectively doubling the number of involved frequency components (see Fig. 1(b)). One could ask what happens in a PT-symmetric arrangement, i.e. if the loss in one ring is exactly balanced by the gain in the other (see Fig. 1(c)).

In the time domain, the interplay between the n[th] longitudinal modes in the two rings obeys a set of two coupled differential equations for their respective modal amplitudes $a_n$, $b_n$:

$$\frac{da_n}{dt} = -i\omega_n a_n + i\kappa_n b_n + \gamma_{a_n} a_n \qquad (1a)$$

$$\frac{db_n}{dt} = -i\omega_n b_n + i\kappa_n a_n + \gamma_{b_n} b_n \qquad (1b)$$

where $\gamma_{a_n} = -\frac{1}{\tau_n} + g_{a_n}$ and $\gamma_{b_n} = -\frac{1}{\tau_n} + g_{b_n}$ represent the net gain/loss in each cavity. The intrinsic losses, including absorption, scattering and radiation, are described by the inverse lifetime $\frac{1}{\tau_n}$ and assumed to be equal for both rings, whereas $g_{a_n}$, $g_{b_n}$ indicate the gain coefficients of the respective resonators. By defining an effective interaction length $L_{eff}$, the coupling $\kappa_n$ between the two rings of radius $R$ and effective index $n_{eff}$ can be related to the coupling coefficient $K_n$ as commonly defined between two parallel waveguides [25] $\kappa = K L_{eff} \frac{c}{2\pi R n_{eff}}$. Assuming an evolution of the form $(a_n, b_n) = (A_n, B_n)e^{-i\omega t}$, the eigenfrequencies $\omega_n^{(1,2)}$ of the two supermodes of this system are given by

$$\omega_n^{(1,2)} = \omega_n + i\frac{\gamma_{a_n} + \gamma_{b_n}}{2} \pm \sqrt{\kappa_n^2 - \left(\frac{\gamma_{a_n} - \gamma_{b_n}}{2}\right)^2}. \qquad (2)$$

Clearly, these eigenfrequencies can in general be complex numbers, and their imaginary part describes amplification or attenuation, respectively. In the PT-symmetric case, i.e. $\gamma_{a_n} = -\gamma_{b_n} \equiv \gamma_n$, Eq. (2) simplifies to

$$\omega_n^{(1,2)} = \omega_n \pm \sqrt{\kappa_n^2 - \gamma_n^2}. \qquad (3)$$

This last relation clearly indicates that the threshold for PT-symmetry-breaking signifies the boundary between amplification/attenuation and bounded neutral oscillations. Any pair of modes, whose gain/loss remains below the coupling coefficient ($\gamma_n(\omega) < \kappa_n(\omega)$), will remain neutral. The absence of any overall gain or loss is easily understood considering that the modes reside equally in the amplifying and lossy regions, as shown in Fig. 2. However, as soon as the gain/loss exceeds the coupling ($\gamma_n(\omega) > \kappa_n(\omega)$), PT symmetry will be broken and a conjugate pair of lasing/decaying modes emerges.

Clearly, a judicious placement of this PT threshold will allow a complete suppression of all non-broken mode pairs in favor of a single amplified mode (see Fig. 1(c)). As the imaginary parts of the eigenvalues diverge from one another, degeneracy between their real parts is restored. We would like to emphasize that in principle any single resonator with a non-uniform gain distribution $g(\omega)$ can exhibit single-mode operation, provided that for all but one resonance the losses overcompensate the gain. However, in this regime, the amplification cannot exceed the gain contrast $g_{max} = g_0 - g_1$ between adjacent resonances (see Fig. 1(a)). Here, $g_0$ refers to the gain of the principal mode, whereas $g_1$ to that of the strongest competing resonance. Obviously, this approach will impose severe constraints on the operating parameters – especially in the case of broad gain windows and/or closely spaced resonator modes, where $g_{max}$ is very small. On the other hand, in the PT-symmetric setting, the coupling $\kappa$ plays the role of an artificial loss. As such, all undesirable modes must fall below its corresponding threshold. According to Eq. (3), we can therefore calculate the maximum achievable gain via $g_{max,PT} = \sqrt{g_0^2 - \kappa^2}$, and by setting $g_1 = \kappa$, we finally arrive at

$$g_{max,PT} = \sqrt{g_0^2 - g_1^2} = g_{max} \cdot \sqrt{\frac{g_0/g_1 + 1}{g_0/g_1 - 1}}. \qquad (4)$$

Evidently, selective breaking of PT symmetry can systematically increase the available amplification during single-mode operation. As a matter of fact, this enhancement is characterized by the factor $G = g_{max,PT}/g_{max}$ and surprisingly becomes more efficient as the gain contrast between adjacent modes decreases (see Fig. 2(c)).

## 3. Experiments

### 3.1 Sample preparation

To experimentally verify our findings, we employed lithographic techniques (see Fig. 3(a)) for the realization of active ring resonators in a layer comprised of six InGaAsP (Indium-Gallium-Arsenide-Phosphide) quantum wells residing on the surface of an InP (Indium Phosphide) substrate [26]. After spin-coating an HSQ (Hydrogen silsesquioxane) resist layer on the wafer, the desired waveguide geometries were patterned by electron beam exposure, allowing for the surrounding InGaAsP to be removed by dry-etching. Figure 3(b) shows an SEM (scanning electron microscope) picture of a typical set of ring structures at this point in fabrication. A $SiO_2$ (silicon dioxide) coating was then applied by means of PECVD (plasma-enhanced chemical vapor deposition). Note that this material blends seamlessly with the $SiO_2$ formed from the electron-exposed HSQ. In a final preparation step, the wafer was bonded to a glass substrate for mechanical support, allowing the remaining InP to be completely removed by wet etching with HCl (Hydrochloric acid). The finished sample therefore consists of InGaAs rings partially buried in $SiO_2$. The gain and loss regions can be defined by selective pumping. Figure 3(c) shows the photoluminescence spectrum obtained when exposing the unstructured quantum well layer to a pump laser at wavelength of 1064 nm. Whereas the active medium itself is capable of providing amplification in a range from 1350 nm to 1600 nm, the net gain in the resonators also depends on the waveguide losses, which generally increase with wavelength and inverse radius of curvature. As a result, smaller resonators tend to exhibit lasing at resonance wavelengths below the photoluminescence maximum of the pristine quantum wells.

### 3.2 Experimental results

The resonators were characterized by homogeneously illuminating them with a pump beam and simultaneously monitoring the intensity distribution and spectrum of the light oscillating in the microrings. To this end, we imaged the sample onto a CCD camera and a spectrometer. Figures 4(a,b) illustrate the behavior of a single ring (radius 5 μm, ring width 300 nm) when exposed to an effective average pump power of 2 mW (15 ns pulses with a repetition rate of 290 kHz). Under these conditions, the signatures of a number of modes can be detected, but two of them dominate the spectrum. When two such rings are placed at 1 μm from one another, and both supplied with the same pump power (Figs. 4 (c,d)), one can clearly see the coupling-induced mode splitting, which occurs symmetrically around the resonance wavelengths of each ring in isolation. In this coupled regime, both structures are contributing equally. As expected, once PT-symmetry is established by withholding the pump from one of the resonators (Figs. 4(e,f)), lasing occurs exclusively in the active ring, where single-mode operation is now achieved. The presence of the lossy ring serves to suppress the unwanted longitudinal modes with a contrast of nearly 30 dB.

Note that no particular constraints in terms of design had to be taken into account when fabricating this double-ring arrangement. The mechanism responsible for mode suppression solely relies on the breaking of PT symmetry, and as such naturally selects the mode with the largest individual gain, even when a single ring would be capable of supporting a number of competing modes. Along these lines, we realized a second set of microring resonators. With a radius of 10 µm, the resonances are more closely spaced. In this case, at least four modes contribute significantly to lasing in the isolated ring (Fig. 5(a,b)). Nevertheless, the transition from evenly pumped double ring (Figs. 5(c,d)) to the PT-symmetric arrangement (Fig. 5(e,f)) reliably enforces single-mode operation.

## 5. Conclusion

In conclusion, we have demonstrated for the first time how the selective breaking of PT-symmetry can be utilized to enforce stable single mode operation in microring laser resonators. In particular, the maximum achievable gain in PT-symmetric ring pairs is systematically enhanced with respect to the onset of undesired lasing in competing modes. Our experiments indicate that this mechanism of mode selectivity is robust with respect to fabrication inaccuracies, and can accommodate active media with wide gain spectra. Moreover, as the occurrence of PT symmetry breaking is exclusively determined by the relation between net gain and coupling, the arrangement is self-adapting: Its functionality does not rely on external parameters or a specific design. While PT-symmetric arrangements can in principle be adopted for any type of laser cavity, they are particularly suited for the control of longitudinal modes in microring resonators, a previously challenging task. Our results may pave the way for a new family of compact laser designs combining the advantages of multimode cavities and stable single-mode emission.

## Acknowledgments

The authors gratefully acknowledge the financial support from NSF (grant ECCS-1128520), and AFOSR (grants FA9550-12-1-0148 and FA9550-14-1-0037). M.H. was supported by the German National Academy of Sciences Leopoldina (grant LPDS 2012-01).

**Figures**

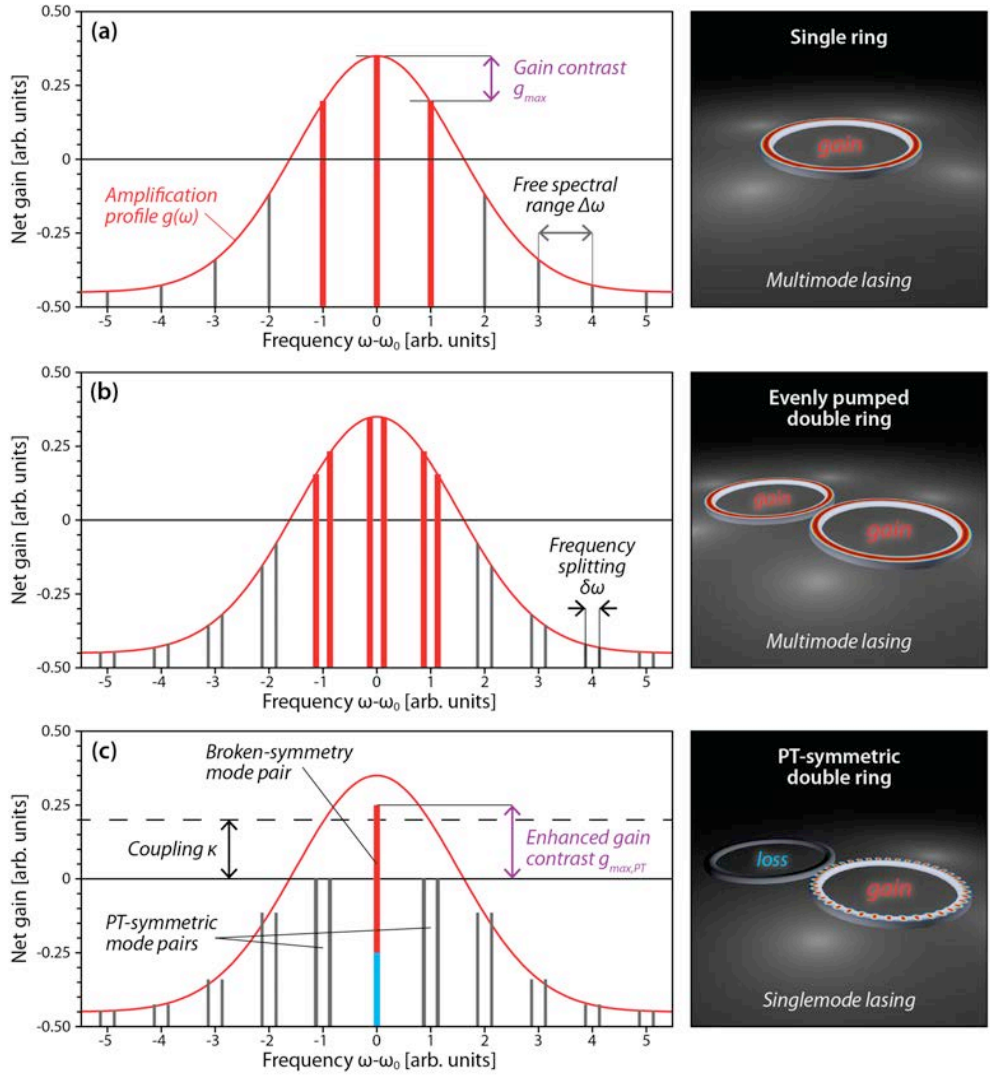

Fig. 1. Schematic principle of mode suppression in PT-symmetric microring lasers. (a) An isolated ring resonator allows lasing of all longitudinal modes with positive net gain. To achieve single-mode operation, maximum permissible gain is limited by the gain contrast between the resonances. (b) In a coupled arrangement of two identical and evenly pumped rings, the degeneracy of resonator modes is broken and mode pairs emerge; their frequency splitting is a measure for the coupling strength. (c) PT-symmetric arrangement: As long as the coupling exceeds the amplification, loss and gain in the two respective rings balance each other, whereas above this threshold, PT symmetry breaking occurs. This mechanism can be exploited to enforce stable single-mode operation in otherwise highly multi-moded resonators. Note that the gain contrast is systematically enhanced with respect to the single-resonator scenario.

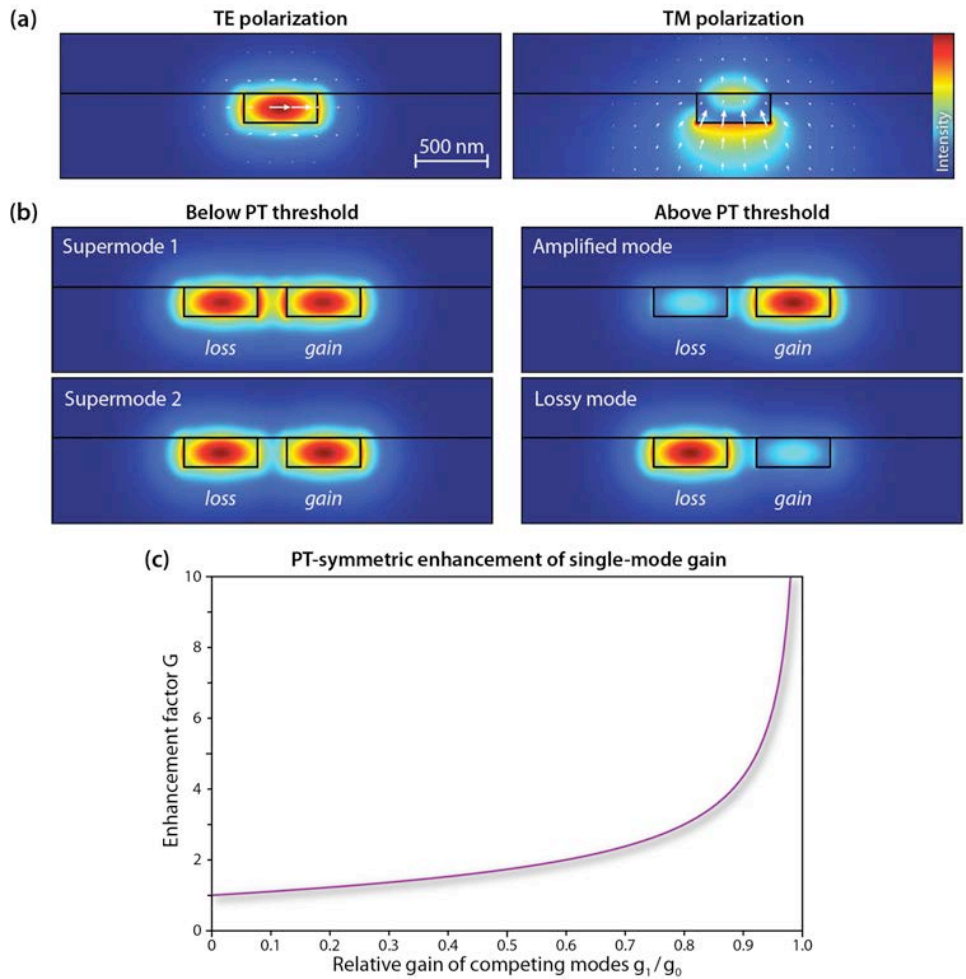

Fig. 2. (a) Transverse intensity distributions of the TE- and TM-polarized modes in a high contrast waveguide such as that used in our experiments. White arrows indicate the local orientation of the electric field. Note that the TM mode resides mostly outside the actual waveguide, and therefore features only a minimal overlap with the active medium. (b) Transverse intensity distributions at the point of closest proximity between the rings. Below the threshold of PT symmetry breaking (left), the supermodes are evenly distributed between the gain and loss regions, respectively. In contrast, above the threshold (right), the modes predominantly reside in one of the rings, and consequently experience a net loss/gain. (c) Enhancement $G$ of the maximum achievable differential gain for single-mode operation in a PT-symmetric setting compared to a single active resonator.

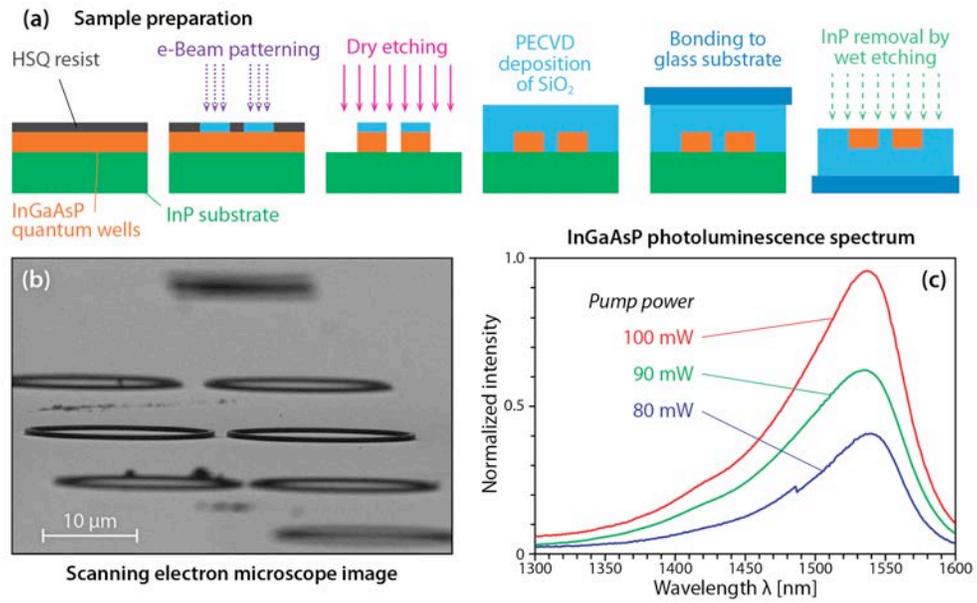

Fig. 3. (a) Schematic of the sample fabrication process. (b) Scanning electron microscope image of a typical set of coupled microring resonator pairs after the dry-etching phase. The exposed structures consist of the InGaAsP quantum well layer and the exposed HSQ resist, which has been transformed into $SiO_2$. (c) Photoluminescence spectrum of the InGaAsP quantum wells as measured for various pump powers in the unstructured layer.

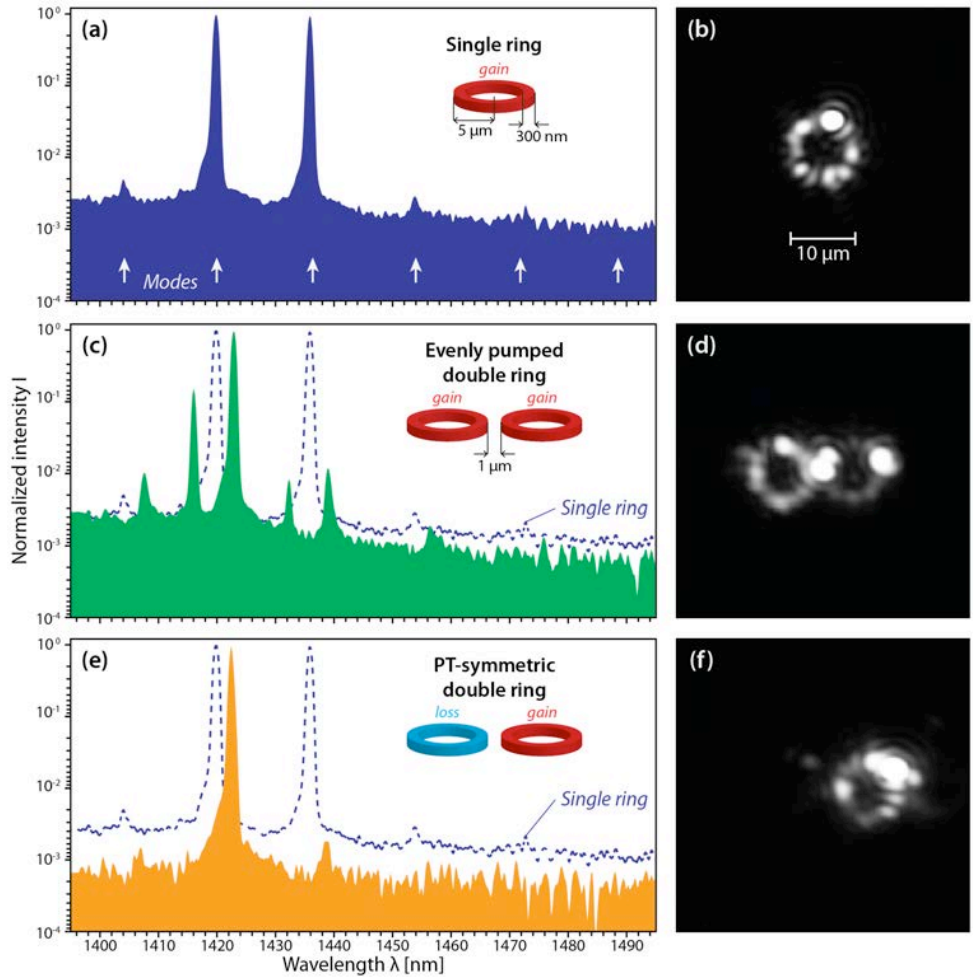

Fig. 4. Experimental observation of mode suppression by PT-symmetry breaking. Case 1: Two competing modes. (a) Emission spectrum of a single resonator: Two modes lase with approximately equal intensity. (b) Corresponding intensity pattern within the ring as observed from scattered light. (c) Spectrum obtained from an evenly pumped pair of such rings (2.0 mW + 2.0 mW). (d) The intensity pattern shows that both resonators are equally involved. (e) Single-moded spectrum under PT-symmetric conditions (0.0 mW + 2.0 mW pump). Coupling to the lossy resonator suppresses the unwanted modes with a contrast of nearly 30dB. (f) Lasing exclusively occurs in the active resonator.

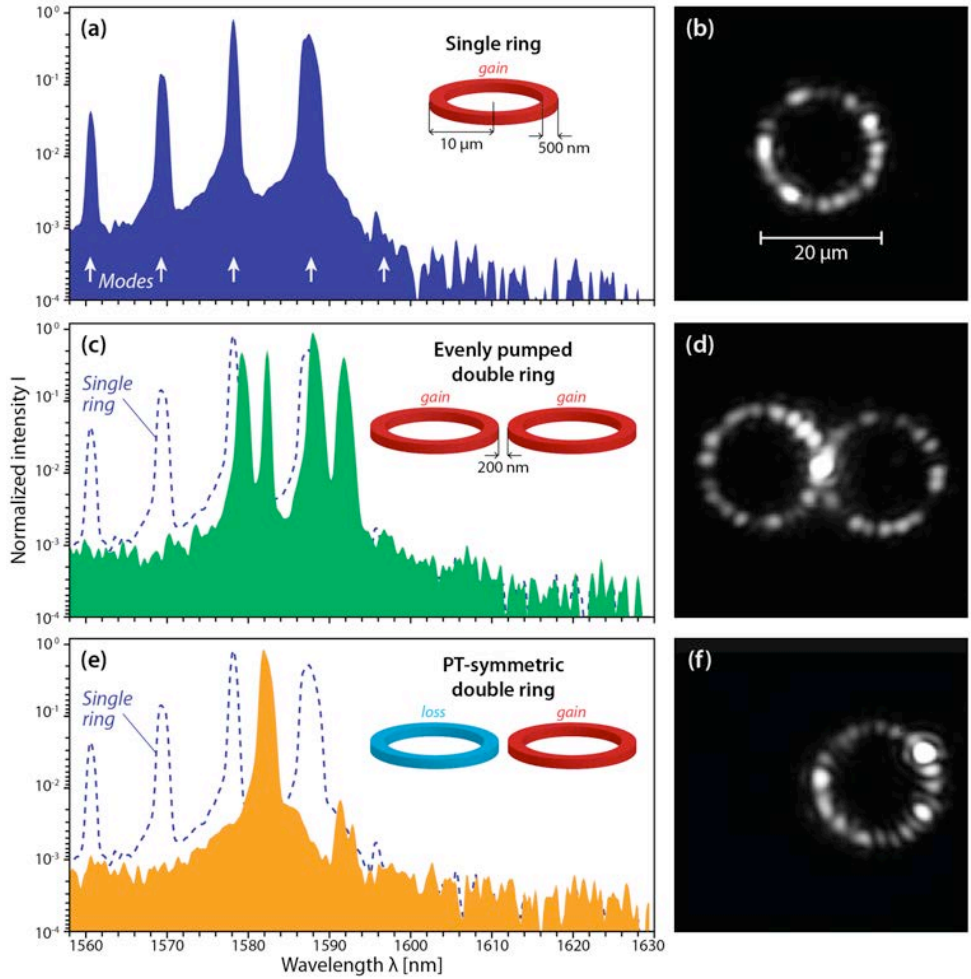

Fig. 5. Experimental observation of mode suppression by PT-symmetry breaking. Case 2: Multiple competing modes. (a) Emission spectrum of a single resonator (radius 10μm, ring width 500 nm) when exposed to an average pump power of 7.4 mW. Due to the larger ring radius, the mode spacing is decreased, and several modes are lasing. (b) Corresponding intensity pattern within the ring as observed from scattered light. (c) Spectrum obtained from an evenly pumped pair of such rings (7.4 mW + 7.4 mW). (d) The intensity pattern shows that both resonators are equally involved. (e) Single-moded spectrum under PT-symmetric conditions (0 mW + 7.4 mW pump). Despite more densely spaced modes of the ring, the mode selectivity remains above 20dB. (f) Lasing exclusively occurs in the active resonator.